# Radiative recombination of charged excitons and multiexcitons in CdSe quantum dots

M. C. Troparevsky* and A. Franceschetti[*]

Oak Ridge National Laboratory, Oak Ridge, TN 37831

*We report semi-empirical pseudopotential calculations of emission spectra of charged excitons and biexcitons in CdSe nanocrystals. We find that the main emission peak of charged multiexcitons - originating from the recombination of an electron in an s-like state with a hole in an s-like state - is blue shifted with respect to the neutral mono exciton. In the case of the negatively charged biexciton, we observe additional emission peaks of lower intensity at higher energy, which we attribute to the recombination of an electron in a p state with a hole in a p state.*

---

[*] Present address: National Renewable Energy Laboratory, Golden, CO 80401

In nanometer-size semiconductor quantum dots multiple electron-hole pairs created by optical excitation can coexist in a confined volume, that is often much smaller than the typical volume occupied by an exciton in bulk semiconductors. As a result, inter-particle (electron-electron, electron-hole, and hole-hole) Coulomb and exchange interactions are strongly enhanced compared to bulk semiconductors, and may affect the spectroscopy of multi excitons and charged excitons[1,2,3]. Unfortunately, the detection of light from multi-exciton recombination in colloidal quantum dots is hindered by the short lifetime of multi-exciton states. Pump-probe experiments[4] have demonstrated that the primary decay channel for multi-particle excitations in colloidal nanocrystals consists of non-radiative Auger recombination of electron-hole pairs. In this energy-conserving process, a multi exciton $X^n$ (where n=2 for a biexciton, n=3 for a triexciton, etc.) decays into an $X^{n-1}$ (multi)exciton by promoting an electron to a higher-energy conduction-band level, or a hole to a lower-energy valence-band level. This process - enabled by the coupling of $X^n$ and $X^{n-1}$ electronic states via a screened Coulomb interaction[5] - has a characteristic lifetime of ~10-100 ps. As a result, the time-integrated photoluminescence signal comes almost entirely from the recombination of single excitons, created either by direct optical excitation, or by ~ps Auger recombination of multi-excitons.

Recently, time-resolved, femtosecond photoluminescence measurements have detected emission from short-lived, multi-particle excited states in CdSe nanocrystals[6,7,8,9], as evidenced by the appearance of additional emission bands when the excitation intensity increases. However, the physical origin of the short-lived emission lines and the underlying optical transitions are still controversial. Achermann et al.[6] measured multiexciton spectra of CdSe/ZnS core/shell nanocrystals. They observed two additional bands at large excitation density, one to the red and one to the blue of the single exciton line. They attributed these bands to the emission from neutral and negatively charged biexciton, respectively. Bonati et al.[8] observed three emission lines in addition to the single exciton line. Two of those bands were redshifted while one was blueshifted with respect to the single exciton peak. The redshifted lines were attributed to the neutral and positively charged biexciton, while the blueshifted line was attributed to the triexciton. Wang et al.[10] reported photoluminescence (PL) spectra of electrically charged CdSe nanocrystals. They found that the PL spectrum of negatively charged nanocrystals (with one electron per dot on average) is nearly identical to that of neutral nanocrystals.

Previous pseudopotential calculations[3] have shown that the *absorption* spectra of charged quantum dots differ significantly from those of neutral dots, as a result of interparticle interactions and state filling. In this work we report pseudopotential calculations of *emission* spectra of charged excitons and biexcitons in CdSe quantum dots. We study the mono-exciton X (one electron, one hole, 1e-1h), the charged excitons $X^-$ (2e-1h) and $X^+$ (1e-2h), and the charged biexcitons $XX^-$ (3e-2h) and $XX^+$ (2e-3h). We find that the lowest-energy emission peak for the $X^-$ recombination is slightly redshifted (by 5 meV) with respect to the mono-exciton, while the $X^+$ emission peak is blue shifted by 15 meV. We also find that the main emission peaks from the $XX^-$ and $XX^+$ recombination are blue shifted by 70 and 89 meV, respectively, with respect to the single exciton X. These blue shifts are due to inter-particle Coulomb and exchange interactions. In the case of $XX^-$, we observe two additional peaks at higher energy that we attribute to the recombination of a p electron with a p hole.

We consider nearly spherical CdSe quantum dots, 3.9 nm in diameter, obtained by cutting out a fragment from a CdSe bulk crystal in the wurtzite lattice structure. The dangling bonds at the surface of the quantum dot are passivated using a ligand-like potential[11] that shifts the energy of the surface states removing them from the band gap. The calculation of the emission spectrum is performed in three steps:

(i) First, we calculate the single-particle energies $\varepsilon_i$ and wave functions $\psi_i$ of the quantum dot using the semi-empirical pseudopotential method (SEPM) described in Refs. 11-[12]. In this approach, one solves for the single-particle Schrödinger equation:

$$[-\nabla^2 + V(\boldsymbol{r}) + \hat{V}_{NL}]\psi_i(\boldsymbol{r}) = \varepsilon_i \psi_i(\boldsymbol{r}) \qquad (1)$$

where $V(\boldsymbol{r}) = \sum_R v(\boldsymbol{r} - \boldsymbol{R})$ is a superposition of screened atomic pseudopotentials (centered at the atomic positions $\boldsymbol{R}$) that have been fitted to bulk experimental transition energies, effective masses, and deformation potentials, and to ab-initio calculated bulk wave functions. $\hat{V}_{NL}$ is a non-local potential that includes spin-orbit interactions[13].

(ii) In the next step, we calculate multi exciton and charged exciton energies $E^{(i)}$ and wave functions using the configuration-interaction method discussed in Ref. [14]. The wave functions $\Psi^{(i)}$ are expanded in a basis set of multi-particles Slater determinants obtained by promoting one or more electrons from the valence band to the conduction band. The Slater determinants are coupled via a configuration-interaction Hamiltonian that includes electron-electron, electron-hole, and hole-hole Coulomb and exchange interactions, which are given by the integrals:

$$J_{ij,kl} = \iint \psi_i^*(\boldsymbol{r})\psi_j^*(\boldsymbol{r}') \frac{e^2}{\bar{\varepsilon}(\boldsymbol{r},\boldsymbol{r}')|\boldsymbol{r}-\boldsymbol{r}'|} \psi_k(\boldsymbol{r})\psi_l(\boldsymbol{r}') d\boldsymbol{r}d\boldsymbol{r}'. \qquad (2)$$

Here $\bar{\varepsilon}(\boldsymbol{r},\boldsymbol{r}')$ is a phenomenological screening function that depends on the size of the quantum dot[14].

(iii) Finally, the emission spectrum is calculated as:

$$\sigma_i(E) = \sum_f \left| \langle \Psi^{(i)} | r | \Psi^{(f)} \rangle \right|^2 g(E - E^{(f)} + E^{(i)}), \qquad (3)$$

where *r* is the dipole operator and *g* is a Gaussian broadening function. Since intra-band carrier relaxation occurs on a much faster time scale than radiative recombination, we assume that the initial states are thermally populated according to a Boltzmann distribution.

The exciton energy levels of CdSe nanocrystals were calculated in the past using the SEPM approach, and were found to be in very good agreement with experiment [11, 13]. For instance, the differences between calculated and experimental optical gaps range between 0.1 eV to 0.2 eV, while the difference in fine-structure splittings, such as the dark-bright exciton splitting, are only a few meV. We expect a similar level of accuracy in our present calculations.

The near band-edge electron and hole single-particle levels of the 3.9 nm CdSe nanocrystal are shown in Table I. The level at the bottom of the conduction band ($e_s$) and the first two levels at the top of the valence band ($h_{s1}$ and $h_{s2}$) have s-like envelope functions. The states $h_{s1}$ and $h_{s2}$ are separated by the crystal field splitting due to the wurtzite lattice structure. The next three conduction-band levels ($e_{p1}$, $e_{p2}$, and $e_{p3}$) and two valence-band levels ($h_{p1}$ and $h_{p2}$) have p-like envelope functions.

Figure 1 shows the calculated emission spectra for $XX^+$, $X^+$, $X$, $X^-$, and $XX^-$ recombination at room temperature (solid lines) and at 5K (dashed lines). Table II summarizes the optical transitions at the origin of the main peaks appearing in Fig. 1. An analysis of Fig.1 and Table II suggests the following observations:

(i) The peaks denoted $A_1$ in Fig.1 and Table II originate from the radiative recombination of an electron in the s-like state $e_s$ with a hole in the s-like state $h_{s1}$. The $A_1$ peaks are the dominant emission peaks in all cases considered here. In the case of the single exciton X, electron-hole exchange interactions split the lowest excitonic manifold into a lower-energy, "dark" exciton and a higher energy, "bright" exciton, separated by ~5 meV. The $A_1$ emission line of the single exciton comes from recombination of the thermally populated bright state, as demonstrated by the low emission intensity at 5 K (see Fig. 1).

(ii) Peaks $A_1$' and $A_1$'' also originate from $h_{s1}$-$e_s$ recombination. Peaks $A_1$' in the $XX^+$ and $XX^-$ spectra (see Fig. 1) can be traced to the transition from the $XX^+$ ($XX^-$) ground state to an excited spin configuration of $X^+$ ($X^-$). These peaks are red shifted by 28 meV (76 meV) with respect to the $A_1$ peak in the $XX^+$ ($XX^-$) spectrum. Peaks $A_1$'' in the $X^+$ and $XX^-$ spectra originate from higher-energy initial states that are thermally populated at room temperature.

(iii) Peaks $A_2$ originates from the recombination of an electron in the $e_s$ state with a hole in the $h_{s2}$ state (see Fig. 1 and Table I). The $h_{s2}$ level is unoccupied in the ground state configuration of the exciton complexes considered here. Thus, the $A_2$ transitions are present only at room temperature but not at low temperature.

(iv) Peaks $B_1$ and $B_1$' in the $XX^-$ emission spectrum originate from the recombination of an electron in the $e_p$ states with a hole in the $h_{p1}$ state. The presence of those peaks is surprising, since in the single-configuration approximation the ground state of $XX^-$ does not have a hole in the $h_{p1}$ state (both holes occupy s-like states). We find that peak $B_1$ (present at low temperature) is due to the fact that the ground state of $XX^-$ has a small but non-negligible mixing with a configuration with a hole in the $h_{p1}$ state. Peak $B_1$', on the other hand, is due to thermal occupation of higher-energy initial states, and occurs only at room temperature.

Wang et al.[14] reported a negligible shift of the emission peak of the negatively charged exciton ($X^-$ in our notation) with respect to the neutral exciton (X) for CdSe nanocrystals ~6.8 nm in diameter. Our calculations are in agreement with the results of Ref. 14: For a 3.9 nm diameter nanocrystal we find that the emission peak ($A_1$) of the negatively charged exciton $X^-$ is redshifted by only 5 meV with respect to the emission peak of the neutral exciton X (see Fig. 1 and Table II). We expect the red shift to be even smaller in larger nanocrystals, such as those studied in Ref. 10. Bonati et al.[8] observed an emission peak to the red of the single-exciton line in CdSe nanocrystals capped by organic ligands. The measured red shift was 152 meV for nanocrystals ~4nm in diameter, and was attributed to emission from positively charged biexcitons (XX+). We do not find evidence of a significant red shift in any of the excitonic complexes considered here, suggesting that the red-shifted emission observed in Ref. 8 may have a different spectroscopic origin. Achermann et al.[14] reported, for a 2.2 nm radius CdSe dot, an emission band 160 meV to the blue of the monoexciton line. They proposed that this emission band could originate from s-p recombination or p-p recombination of negatively charged biexcitons ($XX^-$). We find that the intensity of the s-p recombination is very small in all cases considered here, so s-p recombination is not likely to explain the observed high-energy emission line. We do observe in our calculated $XX^-$ spectrum peaks corresponding to p-p recombination ($B_1$ and $B_1'$ in Fig. 1), but their intensity is rather low, and their energy is higher than the emission peak observed in Ref. 6.

From Fig. 1 and Table II one can see that the $A_1$ peak shifts to higher energy when several spectator electrons and/or holes are present in the quantum dot. For example, in the case of the negatively charged biexciton XX⁻, peak $A_1$ is blue shifted by $\delta E(A_1) = 70$ meV with respect to the monoexciton X. A similar blue shift is predicted for the positively charged biexciton XX⁺ (Fig.1 and Table II). These shifts are due primarily to direct inter-particle interactions (hole-hole, electron-electron, and electron-hole). We can estimate the effects of inter-particle interactions using first order perturbation theory to calculate the energy of the excitonic transitions. In the case of XX⁻, we obtain:

$$\delta E(A_1) = J_{sp}^{ee} + J_{ss}^{ee} + J_{ss}^{hh} - 2J_{ss}^{he} - J_{sp}^{he}, \qquad (4)$$

where $J_{ij}$ are the diagonal Coulomb energies, and we have neglected exchange interactions. Eq. (3) gives a blue shift $\delta E(A_1) = 97$ meV. We have assumed in our calculations that the missing hole in XX⁻ is far removed from the nanocrystal. In reality, it is possible that the hole could be trapped near the surface of the nanocrystal (or near the interface between the CdSe core and the ZnS shell). This would cause a rearrangement of the electron and hole wave functions, as discussed in Ref.[15]. As a result, the electron-electron and hole-hole repulsion, $J^{hh}$ and $J^{ee}$, would increase, while the electron-hole attraction $J^{he}$ would decrease compared to the case where the hole is trapped away from the nanocrystal. One can see from Eq. (3) that the overall effect would be an increase in the value of the shift $\delta E(A_1)$. Given that the Coulomb energies in Eq (1) are of the order of 150-250 meV, we predict that trapping of a hole near the surface of the nanocrystal would produce a significant additional blue shift of the XX⁻ emission peak. This may explain the large blue shift (~160 meV) reported in Ref. 6.

In summary, we have calculated the emission spectra of charged excitons and biexcitons of a CdSe quantum dot. We find that the main peaks for XX⁺ and XX⁻ recombination are significantly blueshifted with respect to the monoexciton peak because of interparticle interactions. The main peak for XX- originates from the recombination of an electron in an s state with a hole in an s state. XX$^-$ also presents two additional emission peaks around 2.5 eV, which we attribute to the recombination of an electron in a p state with a hole in a p state.

This work was supported by the U.S. Department of Energy, Office of Science, Basic Energy Sciences initiative LAB03-17. This research used resources of the National Energy Research Scientific Computing Center, which is supported by the Office of Science of the U.S. Department of Energy under Contract No. DE-AC03-76SF00098.

Table I. Calculated band-edge single-particle levels (in eV) of a 3.9 nm-diameter CdSe nanocrystal.

| Hole state | Energy (eV) |
|:---:|:---:|
| $h_{s1}$ | -5.561 |
| $h_{s2}$ | -5.589 |
| $h_{p1}$ | -5.599 |
| $h_{p2}$ | -5.602 |
| **Electron state** | **Energy (eV)** |
| $e_s$ | -3.211 |
| $e_{p1}$ | -2.912 |
| $e_{p2}$ | -2.904 |
| $e_{p3}$ | -2.901 |

Table II. Origin of main peaks observed in the emission spectra of Fig.1. Column 3 shows the shift of each peak with respect to peak "a" of the absorption spectrum. Column 4 shows the main optical transition responsible for each peak. Boldface characters indicate the single-particle states occupied by the electron and the hole that recombine.

| Exciton | Peak | Shift (meV) | Transition |
|---|---|---|---|
| XX+ | $A_1$ | 89 | $\mathbf{h_{s1}}\, h_{s1}\, h_{p1},\, \mathbf{e_s}\, e_s \rightarrow h_{s1}\, h_{p1},\, e_s$ |
|  | $A_1'$ | 61 | $\mathbf{h_{s1}}\, h_{s1}\, h_{p1},\, \mathbf{e_s}\, e_s \rightarrow h_{s1}\, h_{p1},\, e_s$ |
|  | $A_2$ | 133 | $h_{s1}\, h_{s1}\, \mathbf{h_{s2}},\, \mathbf{e_s}\, e_s \rightarrow h_{s1}\, h_{s1},\, e_s$ |
| X+ | $A_1$ | 40 | $\mathbf{h_{s1}}\, h_{s1},\, \mathbf{e_s} \rightarrow h_{s1}$ |
|  | $A_1'$ | 15 | $\mathbf{h_{s1}}\, h_{p1},\, \mathbf{e_s} \rightarrow h_{p1}$ |
|  | $A_2$ | 79 | $h_{s1}\, \mathbf{h_{s2}},\, \mathbf{e_s} \rightarrow h_{s1}$ |
| X | $A_1$ | 0 | $\mathbf{h_{s1}},\, \mathbf{e_s} \rightarrow 0h;\, 0e$ |
|  | $A_2$ | 33 | $\mathbf{h_{s2}},\, \mathbf{e_s} \rightarrow 0h;\, 0e$ |
| X- | $A_1$ | -5 | $\mathbf{h_{s1}},\, \mathbf{e_s}\, e_s \rightarrow e_s$ |
|  | $A_2$ | 33 | $\mathbf{h_{s2}},\, \mathbf{e_s}\, e_s \rightarrow e_s$ |
| XX- | $A_1$ | 70 | $\mathbf{h_{s1}}\, h_{s1},\, \mathbf{e_s}\, e_s\, e_p \rightarrow h_{s1},\, e_s\, e_p$ |
|  | $A_1'$ | -6 | $\mathbf{h_{s1}}\, h_{s1},\, \mathbf{e_s}\, e_s\, e_p \rightarrow h_{s1},\, e_s\, e_p$ |
|  | $A_1'$ | 36 | $\mathbf{h_{s1}}\, h_{p1},\, \mathbf{e_s}\, e_s\, e_p \rightarrow h_{p1},\, e_s\, e_p$ |
|  | $A_2$ | 107 | $h_{s1}\, \mathbf{h_{s2}},\, \mathbf{e_s}\, e_s\, e_p \rightarrow h_{s1},\, e_s\, e_p$ |
|  | $B_1$ | 294 | $\mathbf{h_{p1}}\, h_{p1},\, e_s\, e_s\, \mathbf{e_p} \rightarrow h_{p1},\, e_s\, e_s$ |
|  | $B_1'$ | 369 | $h_{s1}\, \mathbf{h_{p1}},\, e_s\, e_s\, \mathbf{e_p} \rightarrow h_{s1},\, e_s\, e_s$ |

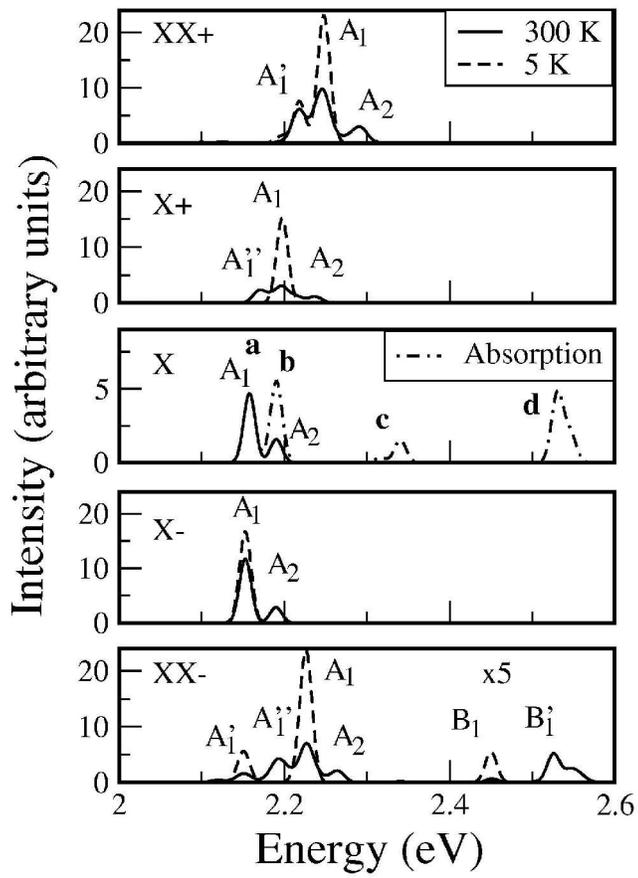

**Fig. 1**

**Figure captions**

Fig. 1. Calculated emission spectra of $XX^+$, $X^+$, $X$, $X^-$, and $XX^-$ at room temperature (solid lines) and at 5 K (dashed lines). The center plot (X) also shows the calculated, low-temperature absorption spectrum (dashed line). "A" peaks originate from recombination of an electron in an s-like state with a hole in an s-like-state, while "B" peaks originate from the recombination of a p electron with a p hole (see text). For clarity purposes, the room temperature spectra have been shifted rigidly on the x axis to align the $A_1$ peaks with those of the low temperature spectra. All spectra were broadened using a Gaussian with a 0.01 eV width.